\documentclass[traditabstract]{aa} 
\usepackage{graphicx}
\usepackage{lscape}
\usepackage{natbib}
\usepackage{txfonts}
\usepackage{longtable}
\begin{document}
\newcommand{\kms}{km~s$^{-1}$}
\newcommand{\Msun}{M$_{\odot}$}
\newcommand{\Teff}{$T_{\rm eff}$}
\newcommand{\FeH}{[Fe/H]}
\newcommand{\Cir}{$^{12}$C / $^{13}$C}
\newcommand{\bacchus}{{\footnotesize BACCHUS}}
\newcommand{\vald}{{\footnotesize VALD}}
\title {IP Eri: A surprising long-period binary system hosting a He white dwarf}

   \author{T. Merle
          \inst{1}
          \and A. Jorissen
          \inst{1}
          \and T. Masseron
          \inst{1}
          \and S. Van Eck
          \inst{1}
          \and L. Siess
          \inst{1}
          \and H. Van Winckel
          \inst{2}
  }

\institute{Institut d'Astronomie et d'Astrophysique, Universit\'e Libre de Bruxelles, CP. 226, Boulevard du Triomphe, 1050 Brussels, Belgium \\
              \email{tmerle@ulb.ac.be}
\and Instituut voor Sterrenkunde, Katholieke Universiteit Leuven, Celestijnenlaan 200D, 3200 Heverlee, Belgium}

\date{Received ...; accepted ...}

 \abstract
   {We determine the orbital elements for the K0~IV + white dwarf (WD)  system IP Eri, which appears to have a surprisingly long period of 1071~d and a significant eccentricity of 0.25. Previous spectroscopic analyses of the WD, based on a distance  of 101~pc inferred from its Hipparcos parallax, yielded a mass of only 0.43~\Msun, implying it to be a helium-core WD. The orbital properties of IP~Eri are similar to those of the newly discovered long-period subdwarf B star (sdB) binaries, which involve stars with He-burning cores surrounded by extremely thin H envelopes, and are therefore close relatives to He WDs. We performed a spectroscopic analysis of high-resolution spectra from the HERMES/Mercator spectrograph and concluded that the atmospheric parameters of the K0 component are \Teff$=4960$~K, $\log{g}=3.3$, [Fe/H$] = 0.09$ and $\xi=1.5$~\kms. The detailed abundance analysis focuses on C, N, O abundances, carbon isotopic ratio, light (Na, Mg, Al, Si, Ca, Ti) and \textit{s}-process (Sr, Y, Zr, Ba, La, Ce, Nd) elements. We conclude that IP~Eri abundances agree with those of  normal field stars of the same metallicity. The long period and non-null eccentricity indicate that this system cannot be the end product of a  common-envelope phase; it calls instead for another less catastrophic binary-evolution channel presented in  detail in a companion paper.}

\keywords{Stars: abundances -- binaries: spectroscopic -- Stars: evolution --  Stars: individual: IP Eri -- subdwarfs -- white dwarfs}

\titlerunning{IP Eri}

\maketitle
%

\section{Introduction}

IP~Eri is a very interesting system consisting of a K0 (sub)giant and a He white dwarf (WD) and it is tempting to relate this system to the family of subdwarf B (sdB) binaries where the hot component is a bare He-burning core surrounded by an extremely thin H envelope \citep{Heber2009}. These systems owe their properties to envelope ejection, likely due to binary interaction, as they evolve along the red giant branch. Their evolution shares some similarities with that of IP~Eri and such systems are important benchmarks for binary evolution.

Several long-period eccentric systems (with $P\sim10^3$~d) were recently discovered among sdB stars \citep{Ostensen2011,Ostensen2012,Vos2012,Vos2013,Deca2012,Barlow2012,Barlow2013}. Using new radial-velocity data collected with the HERMES/Mercator spectrograph \citep{Raskin2011}, we show in this paper (Sect.~\ref{Sect:orbit}) that IP~Eri adds to this new class of long-period eccentric systems.

Since the He-WD progenitor did not evolve along the asymptotic giant branch (AGB), it had no chance to produce \textit{s}-process elements and to pollute its companion (the present K0 subgiant), so that the latter should not appear as a barium star \citep[a family of K giants with enhanced abundances of \textit{s}-process elements;][]{Bidelman-Keenan-51}. It is therefore of interest to perform a chemical analysis of the K0 subgiant in the IP~Eri system to confirm the absence of overabundances of \textit{s}-process elements. This is the second objective of the present paper, which is organized as follows: Sect.~\ref{Sect:IPEri} gives an overview of the properties of the IP~Eri system. Sect.~\ref{Sect:orbit} presents the radial-velocity data and the ensuing orbit. After deriving the atmospheric parameters  of IP~Eri in Sect.~\ref{Sect:atmosphere}, the abundance analysis is presented in Sect.~\ref{Sect:abundances}, with emphasis on \textit{s}-process elements. These abundances are then compared with expectations for barium stars and for non-\textit{s}-process-polluted stars,  as derived from the abundance trends observed in large samples of field stars reflecting the chemical evolution of the Galaxy  (Sect.~\ref{Sect:evolgal}). Sect.~\ref{Sect:analysis} briefly  confronts our results  with predictions from binary-evolution scenarios presented in  detail in a companion paper \citep{Siess2014}. Sect.~\ref{Sect:summary} summarizes our results.

\section{The IP Eri system}
\label{Sect:IPEri}

IP Eri (HD~18131, HIP~13558, WD 0252-055, EUVE J0254-053) has attracted attention since its discovery as an extreme UV (EUV) source both by \textit{ROSAT} \citep{Pounds1993,Pye1995} and \textit{EUVE} \citep{Bowyer1994,Malina1994,Bowyer1996}. In a subsequent analysis, \citet{Vennes1995} find that an older {\it International Ultraviolet Explorer} (\textit{IUE}) spectrum already revealed that a WD dominates the \textit{IUE} spectrum below 200~nm while a K0 spectrum prevails at longer wavelengths. Their model-atmosphere analysis of the EUV photometry reveals a hot, hydrogen-rich (DA) WD (with an effective temperature of about 30~000~K) that is the most likely source of the \textit{EUV} emission. In the grid of models fitted to the WD spectrum by \citet{Burleigh1997}, the model with a temperature of 29~290~K, a gravity of $\log g = 7.5$ and a mass of 0.43~\Msun\ locates the WD at a distance of 93~pc, consistent with the Hipparcos parallax of the K star, as discussed below. The corresponding age for the WD is then 7~Myr \citep{Burleigh1997}. The WD parameters inferred by \citet{Burleigh1997} locate it among He WDs in the $\log g - \log T_{\rm eff}$ diagram of \citet{Driebe1998}. Its mass of 0.43~\Msun\ is incompatible with a CO WD since the minimum CO core mass at the base of the AGB (just at the end of core He-burning) is 0.51~\Msun\ \citep[for a star of initial mass 0.9~\Msun; e.g., Eq. 66 of ][]{Hurley2000}. An independent study by \citet{Vennes1998} concludes at a somewhat higher mass (0.48 --0.52~M$_{\odot}$) for the IP~Eri WD, which is still, for the most part of this range, compatible with the He nature of the WD.

\begin{figure}
\includegraphics[width=\linewidth]{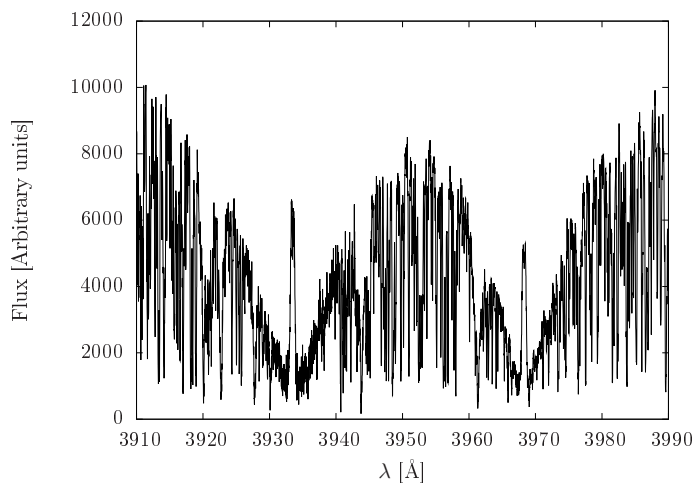}
\caption[]{\label{Fig:CaII}
The \ion{Ca}{II} H and K lines, from the HERMES spectrum on HJD~2455066.748 (2009, August 23). The emission present in the cores is somewhat more intense than the one presented in Fig.~5 of \citet{Vennes1997}.}
\end{figure}

The revised Hipparcos parallax of $9.82\pm0.94$~mas corresponds to a distance of $101\pm11$~pc and a distance modulus of $5.03\pm0.2$ \citep{vanLeeuwen2007}. With $V = 7.32$ \citep{Cutispoto1995}, one gets an absolute magnitude of 2.29 for the K star, which indicates that it is a subgiant. The photometric data for the K0 subgiant are $V = 7.32, U-B = 0.74, B-V = 0.98, V-Rc = 0.52, V-Ic = 0.97$ \citep{Cutispoto1995} and $J = 5.709, H = 5.263, K = 5.090$ \citep[2MASS;][]{Cutri2003}. The corresponding $V - K$ index of 2.23 implies an effective temperature of 4900~K \citep{Bessell-98}. Using a Bayesian method, \citet{Bailer-Jones2011} obtains $\log T_{\rm eff} = 3.70\pm0.01$ (5012~K) for IP~Eri.

Although neither the ASAS lightcurve \citep{Pojmanski1997} nor the monitoring performed by \citet{Cutispoto1999} reveals variability at the 0.1~mag level, a more accurate monitoring by \citet{Strassmeier2000} has uncovered a 0.045~mag variability over 30~d in the Str\"{o}mgren $y$ band, hence its classification as a BY~Dra variable in the General Catalogue of Variable Stars (with name IP~Eri). The star is slowly rotating \citep[$V \sin i < 5$~\kms;][]{Cutispoto1999}. Should the 30~d photometric variability be due to rotation, a value of $\sim 1$~\kms\ for the rotational velocity would result from the radius of 3.8~$R_\odot$  derived in \citep{Siess2014} for IP~Eri. Nevertheless, this star exhibits moderate Mg~II~h and k emission in the \textit{IUE} spectrum and \ion{Ca}{II} H \& K emission \citep[Fig.~5 of][]{Vennes1997,Strassmeier2000}.  Fig.~\ref{Fig:CaII} shows the emission cores in the \ion{Ca}{II} lines, as seen on our HERMES/Mercator spectrum obtained on HJD~2455066.748 (2009, August 23). This emission core is somewhat more intense than the one presented by \citet{Vennes1997}.

\section{Orbital elements}
\label{Sect:orbit}

The 18 high-resolution spectra used to compute the spectroscopic orbital elements were obtained with the HERMES/Mercator spectrograph \citep{Raskin2011}, operating at an average resolution of 85~000 in high-resolution mode and with a spectral range of [$\lambda370 -900$ nm]. The spectra were reduced with the HERMES pipeline, and the radial velocities, computed by cross-correlating the observed spectra with an Arcturus template, are on the IAU wavelength system defined by the standards from \citet{Udry1999a,Udry1999b}. The individual radial velocities $V_r$ are listed in Table~\ref{Tab:Vr}. The errors are dominated by the drift of the air refractive index caused by the atmopsheric pressure variations in the spectrograph room (see Fig.~9 in \citealt{Raskin2011}). The long-term stability during the 4 years of operations of the HERMES spectrograph turns out to be $\sim40$~m~s$^{-1}$ as derived from the standard deviation of the radial velocities of the monitored IAU standards. This may be considered as the precision on the radial velocities produced by HERMES.

\begin{figure}
\includegraphics[angle=-90,width=9cm]{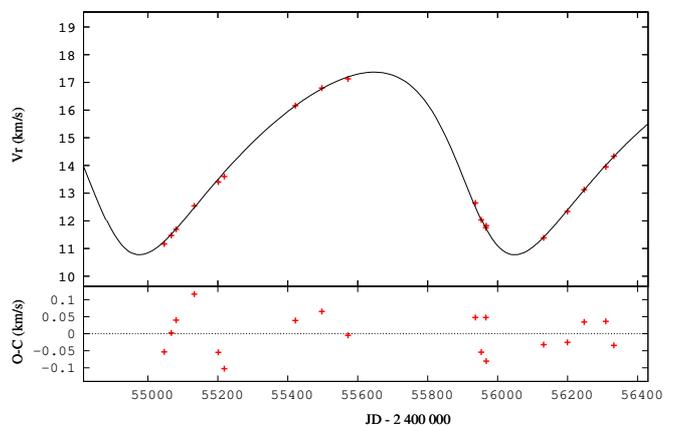}
\caption[]{\label{Fig:orbit}
The orbital solution for IP~Eri. The lower panel shows the observed (O) minus the calculated (C) orbits.}
\end{figure}

The corresponding orbital solution is listed in Table~\ref{Tab:orbit} and displayed in Fig.~\ref{Fig:orbit}. The uncertainties of about 30~m~s$^{-1}$ on the orbital parameters $V_\gamma$ and $K_1$ are consistent with the precision of the spectrograph. Our orbital solution does not include the radial-velocity measurement $V_r =14.94\pm0.10$~km~s$^{-1}$ obtained by  \citet{Chubak2012} on JD 2~455~261, but we checked {\it a posteriori} that it falls on the orbital solution within 0.1~km~s$^{-1}$. It is worth mentioning that IP~Eri was imaged using the \textit{HST} by \citet{Barstow2001} and the binary was not resolved, setting an upper limit on the orbital period of about 19~yr, consistent with the 2.9-yr period found here. No orbit could be adjusted to the O$-$C residuals of the IP~Eri system  (see lower panel of Fig.~\ref{Fig:orbit}) when testing for the presence of a hypothetical third component.

The current mass function  of the system $f\left(M_1,M_2\right) = 0.0036$~M$_{\odot}$ constrains the primary mass $M_1$ to be lower than 4.27~M$_{\odot}$, if
$M_2 = 0.43$~M$_{\odot}$ for the He-WD (see Sect.~\ref{Sect:IPEri}).

\begin{table}
\caption[]{\label{Tab:Vr}
Radial velocities used for computing the orbit of the IP~Eri system. Uncertainties on the radial velocities are about $40$~m~s$^{-1}$.}
\begin{center}
\begin{tabular}{crcl}
\hline\\
HJD & $V_r$~(\kms)\\
\hline\\
2455046.711 & 11.17 \\
2455066.748 & 11.47 \\
2455080.698 & 11.70 \\
2455132.581 & 12.54 \\
2455201.387 & 13.40 \\
2455218.405 & 13.60 \\
2455421.686 & 16.16 \\
2455497.538 & 16.80 \\
2455572.383 & 17.13 \\
2455936.430 & 12.65 \\
2455953.412 & 12.04 \\
2455966.392 & 11.75 \\
2455967.331 & 11.83 \\
2456131.726 & 11.39 \\
2456199.603 & 12.34 \\
2456247.567 & 13.13 \\
2456309.336 & 13.95 \\
2456332.371 & 14.33 \\
\hline\\
\end{tabular}
\end{center}
\end{table}

\begin{table}
\caption[]{\label{Tab:orbit}
Spectroscopic orbital elements of IP Eri.
}
\begin{center}
\begin{tabular}{lrcl}
\hline\\
$\omega\; (^\circ)$  & 128.3             & $\pm$ &  2.0   \\
$e$                  &  0.25             & $\pm$ &  0.01 \\
$P$ (d)              & 1071.0            & $\pm$ &  1.8   \\
$T_0$ (JD)           & $2~455~956.9$ & $\pm$ &  4.7   \\
$V_{\gamma}$ (\kms)  &          +14.59   & $\pm$ &  0.03  \\
$K_1$ (\kms)         &        3.30       & $\pm$ &  0.03  \\
$f(M)$ (M$_{\odot}$) & 0.0036            & $\pm$ & 0.0001 \\

$a_1 \sin i$ (Gm)    & 47.13 \\
$N$                  & 18 \\ 
$\sigma$(O-C) (\kms) & 0.06 \\ 
\hline
\end{tabular}
\end{center}
\end{table}

\begin{figure*}
\begin{center}
 \includegraphics[width=0.8\linewidth]{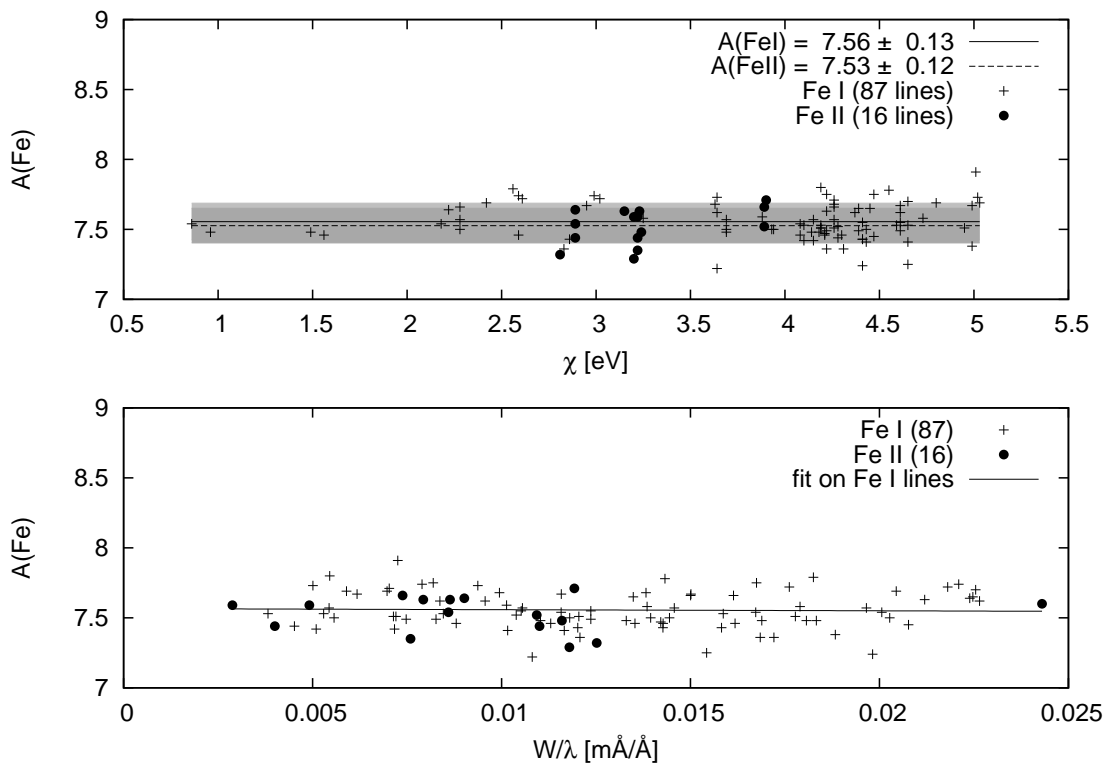}
 \caption{Line by line abundance analysis for \ion{Fe}{i} and \ion{Fe}{ii} lines as a function of the excitation potential $\chi$ and the reduced equivalent width $W/\lambda$. The grey areas represent the standard deviations around the mean abundances of \ion{Fe}{i} and \ion{Fe}{ii}.}
 \end{center}
 \label{fig:fe_abu}
\end{figure*}

\section{Atmospheric parameters}
\label{Sect:atmosphere}

To derive the atmospheric parameters and the detailed abundances, we used two spectra with high signal-to-noise ratios from the list of Table~\ref{Tab:Vr} (namely those obtained on August 22 and September 5, 2009, respectively HJD 2455066.748 and 2455080.698)\footnote{HERMES Spectra are available on electronic form at the CDS via anonymous ftp to \url{cdsarc.u-strasbg.fr} (130.79.128.5) or via \url{http://cdsweb.u-strasbg.fr/cgi-bin/qcat?J/A+A/}}.  The first guesses for the atmospheric parameters were determined from the photometry. The color indices $J-K=0.62$ and $V-K=2.23$ yield a first estimate of \Teff$=4900$~K for the effective temperature, using the calibrations of \citet{Bessell-98}, in perfect agreement with previous estimates. From the calibration of MK spectral types provided by \citet{Cox2000}, we adopted $\log{g}=2.1$ as initial guess for the surface gravity of a giant star of
spectral type K0 (despite the fact that the luminosity class suggests a higher gravity). We initially assumed the metallicity to be solar.

The atmospheric parameters were then determined iteratively using the \bacchus\ pipeline developed by one of the author (TMa; see also \citealt{Jofre-etal-13}) in the context of the Gaia-ESO survey \citep[][]{Gilmore2012}. This pipeline is based on the 1D LTE spectrum-synthesis code Turbospectrum \citep{Alvarez1998,Plez2012} and allows an automated determination of effective temperature \Teff, surface gravity $\log{g}$, metallicity \FeH\ and microturbulent velocity $\xi$. We used MARCS model atmospheres \citep{Gustafsson2008} along with a selection of neutral and singly ionized Fe lines that have been selected for the analysis of  stellar spectra in the framework of the Gaia-ESO survey. Oscillator strengths are from  the VALD database \citep{Kupka00}.  The classical method to obtain the atmospheric parameters consists in avoiding trends in the \FeH\ vs. $\chi$ and \FeH\ vs. $W/\lambda$ relations (where $\chi$ is the lower excitation energy of the considered line, $W$ its measured equivalent width and $\lambda$ its wavelength) but also in forcing lines of \ion{Fe}{I} and \ion{Fe}{II} to yield the same abundance (see Fig.~\ref{fig:fe_abu}).

The equivalent widths are automatically measured using spectrum synthesis with the atmospheric parameters determined at the previous iteration. The synthetic spectra were convolved with a Gaussian function with full width at half maximum of 6.5~\kms. Only iron lines having reduced equivalent widths ($W/\lambda$) lower than 0.025~m\AA/\AA\ were kept in the analysis. We thus obtain the following atmospheric parameters: \Teff$=4960\pm 100$~K, $\log{g}=3.3\pm0.3$, \FeH$=+0.09\pm0.08$ and $\xi=1.5\pm0.1$~\kms. We used the solar reference values from \citet{Grevesse07} where $A_\odot(\mathrm{Fe}) = 7.45$.

\section{Abundances}
\label{Sect:abundances}

The detailed  abundance analysis was performed (in the framework of Local Thermodynamical Equilibrium -- LTE) using the abundance module of the \bacchus\ pipeline. The selection of atomic and molecular lines was performed over the whole wavelength range covered by the HERMES/Mercator spectrograph. The atomic line list used for the detailed analysis is given in Appendix~\ref{ap:ll}. It includes the isotopic shifts for \ion{Ba}{ii} (with an update for isotopes 130 and 132) and the hyperfine structure for \ion{La}{ii} from \citet{Masseron06}. The CH molecular line list is from \citet{Masseron14}. The references for the other molecular line lists (TiO, SiO, VO, C$_2$, CN, NH, OH, MgH, SiH, CaH and FeH) can be found in \citet{Gustafsson2008}. Line fitting is essentially based on a least-square minimization method and all lines are visually inspected to check for possible bad fits (due to, e.g., line blends, cosmic hits, ...). The results of the detailed abundance analysis are presented in Table~\ref{tab:abu}.

\begin{figure*}
\includegraphics[angle=-90, width=\linewidth]{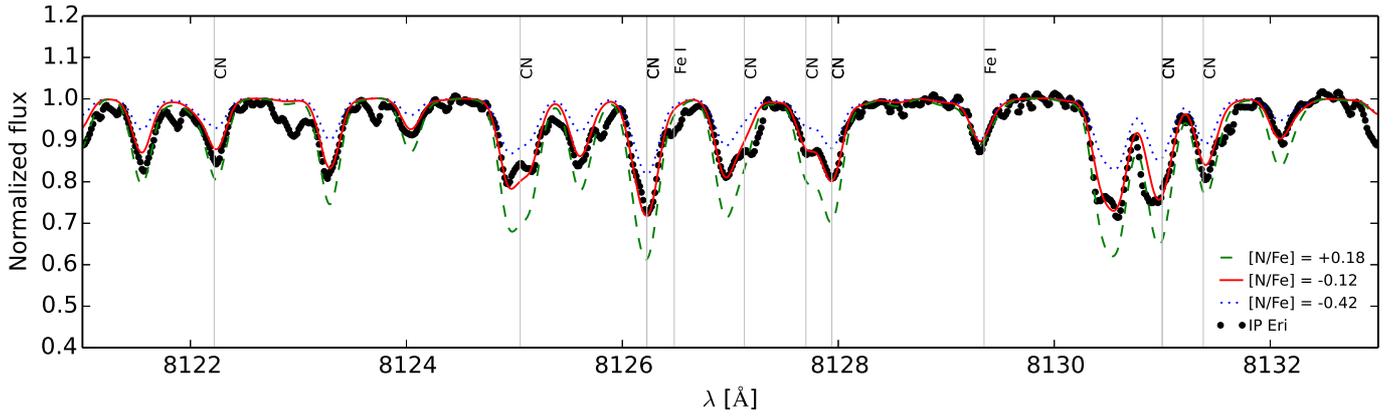}
\caption{Example of CN-line fitting.}
\label{fig:cn_spec}
\end{figure*}
\begin{figure*}
\includegraphics[angle=-90, width=\linewidth]{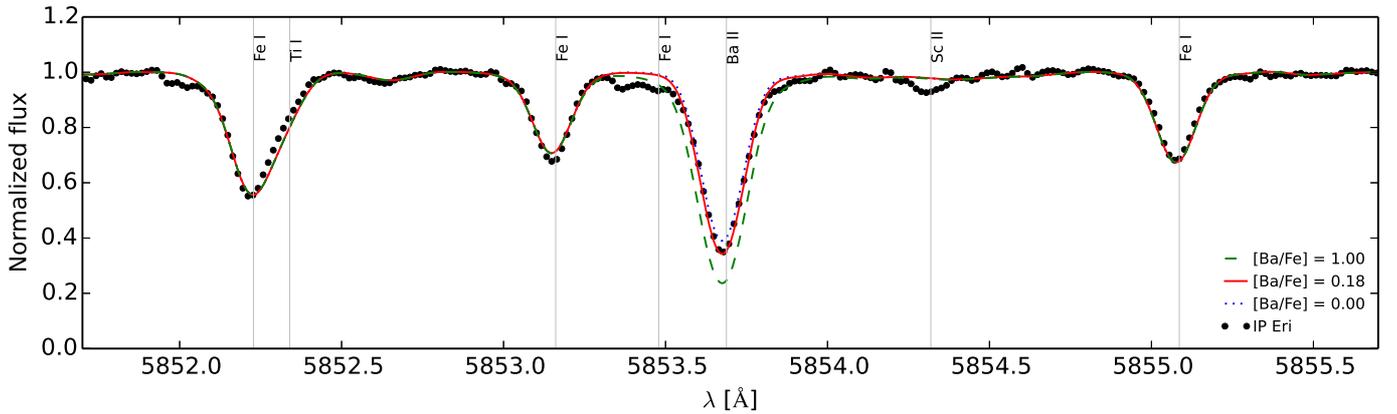}
\caption{Example of \ion{Ba}{ii} line fitting (\ion{Ba}{ii} $\lambda$585.37 nm).}
\label{fig:ba_spec}
\end{figure*}
\begin{figure*}
\includegraphics[angle=-90, width=\linewidth]{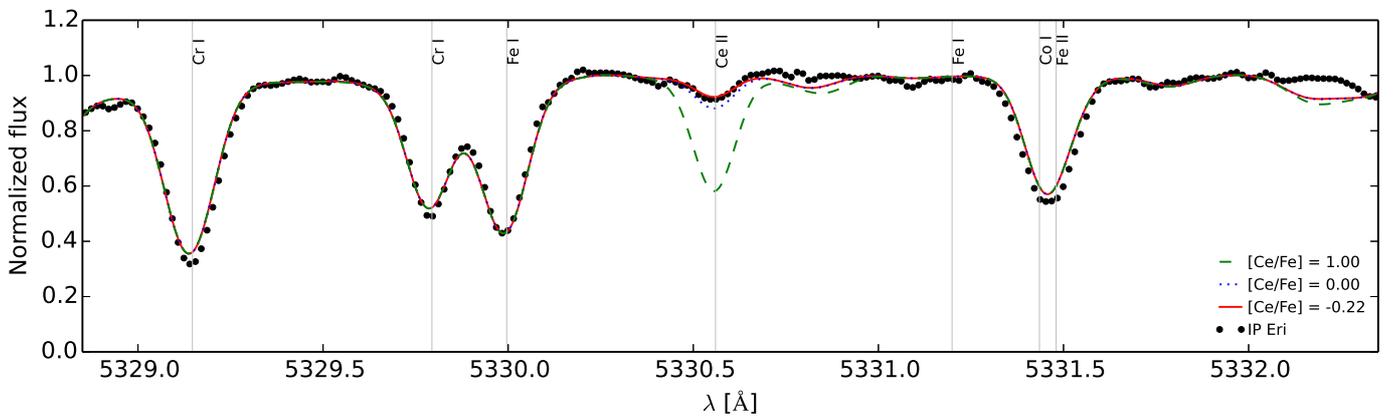}
\caption{Example of \ion{Ce}{ii} line fitting (\ion{Ce}{ii} $\lambda$533.01 nm).}
\label{fig:ce_spec}
\end{figure*}

\begin{table}
 \caption{Results of the chemical abundance analysis. $A({\rm X})$ is the abundance of species X in the logarithmic scale where $A({\rm H}) = 12$.
$\sigma_\mathrm{stat}$ is the line-to-line abundance dispersion. $N$ is the number of lines used for the corresponding species. In the column labelled 'comments', wavelengths are expressed in nm.
}
\tiny
 \begin{tabular}{llcrrcrl}
  X & & A(X) & [X/H] & [X/Fe] & $\sigma_\mathrm{stat}$ &  $N$ & comments\\
  \hline \\
  C & I  & 8.56 & $ 0.17$ & $ 0.08$ & 0.16 & 17 & atomic lines \\
  N &    & 7.75 & $-0.03$ & $-0.12$ & 0.11 &447 & CN lines \\
  O & I  & 8.82 & $ 0.16$ & $ 0.07$ & -    &  1 & $\lambda$630.03 \\
  \\
  Na& I  & 6.45 & $ 0.28$ & $ 0.19$ & 0.11 &  8 & \\
  Mg& I  & 7.80 & $ 0.27$ & $ 0.18$ & 0.10 & 10 & \\
  Al& I  & 6.67 & $ 0.30$ & $ 0.21$ & 0.04 &  7 & \\
  Si& I  & 7.55 & $ 0.04$ & $-0.05$ & 0.10 & 15 & \\
  Ca& I  & 6.59 & $ 0.28$ & $ 0.19$ & 0.04 &  8 & \\
  Ti& I  & 5.14 & $ 0.24$ & $ 0.15$ & 0.15 &  9 & \\
  Ti& II & 5.03 & $ 0.13$ & $ 0.04$ & 0.15 &  5 & \\
  \\ 
  Fe& I  & 7.56 & $0.11$ &         &  $0.13$ &  87 \\
  Fe& II & 7.53 & $0.08$ &         &  $0.12$ &  16 \\
  \\
  Sr& I  & 3.10 & $ 0.18$ & $ 0.09$ & 0.04 &   3 &    \\
  Y & I  & 2.33 & $ 0.12$ & $ 0.03$ & 0.05 &   2 & $\lambda$619.17, $\lambda$643.50 \\
  Y &II  & 1.94 & $-0.27$ & $-0.18$ & 0.11 &   5 &  \\
  Zr& I  & 2.89 & $ 0.31$ & $ 0.22$ & 0.18 &   4 & \\
  Zr& II & 2.82 & $ 0.24$ & $ 0.15$ & 0.01 &   2 & $\lambda$444.30, $\lambda$535.01 \\
  \\
  Ba& II & 2.44 & $ 0.27$ & $ 0.18$ & 0.04 &   4 & \\
  La& II & 1.17 & $ 0.04$ & $-0.05$ & 0.14 &  10 & \\
  Ce& II & 1.57 & $-0.13$ & $-0.22$ & 0.08 &   4 & \\
  Nd& II & 1.50 & $+0.05$ & $-0.04$ & 0.12 &   8 & \\
  \hline
 \end{tabular}
 \normalsize
  \label{tab:abu}
\end{table}

\subsection{C, N, O and \Cir} 
The numerous neutral carbon atomic lines lead to an abundance of [C/Fe$] = 0.08 \pm 0.16$~dex. The nitrogen abundance is determined from numerous molecular CN lines selected over a large spectral range [$\lambda$640 $-$ 890~nm]. An example of the fit of CN lines in the [$\lambda$812.2$-$813.2~nm] region is shown in Fig.~\ref{fig:cn_spec}: synthetic spectra with [N/Fe$]=-0.12\pm 0.3$~dex (see Table~\ref{tab:abu}) are compared with the observed spectrum of IP~Eri. The oxygen abundance is derived from the sole [\ion{O}{I}] $\lambda$630.03~nm line which is supposed to be free from NLTE effects \citep{Asplund2005}. Another forbidden line at $\lambda$636.38~nm is  in the red wing of a \ion{Ca}{I} autoionization line which is difficult to fit in our spectrum. The oxygen triplet at $\lambda$777.19, 777.42 and 777.54~nm gives an abundance $A(\mathrm{O}) \approx 9.46\pm 0.07$~dex, in disagreement by about $0.7$~dex with respect to the $\lambda$630.03~nm line. This discrepancy is mainly due to the NLTE effect of diffusion in the triplet as clearly explained by \citet{Asplund2005}. No OH lines are available in the HERMES spectrum to better constrain the oxygen abundance. The resulting C/O ratio is 0.55, in agreement with the solar value.

The carbon isotopic ratio \Cir\ is deduced from the $^{12}$CN and $^{13}$CN molecular lines in the [$\lambda$799.5--801.5~nm] range \citep[see, e.g.,][]{Barbuy1992,Drake2008}. Specifically, we can compare the $^{12}$CN triplet between $\lambda$800.3 and $\lambda$800.4~nm with the $^{13}$CN feature at $\lambda$800.45~nm, and the $^{12}$CN weak line  at $\lambda$801.0~nm with the $^{13}$CN very weak line at $\lambda$801.05~nm. The latter is in the far red wing of the former. The high resolution of the HERMES spectrograph is able to separate the two components. The best fit of the entire [$\lambda$799.5--801.5~nm] spectral range gives a carbon abundance of $A(\mathrm{C})=8.56\pm 0.10$~dex, which confirms the abundance derived from atomic carbon lines. With this value, we can only deduce a lower limit for the carbon isotopic ratio of \Cir\ $\ge 20$, in accordance with the weakness of the $^{13}$CN features.

\subsection{Light elements}
Neutral lines from Na, Mg, Al, Si and Ca provide reliable abundances with a standard deviation lower than, or of the order of,  0.1~dex. Ca is the most enriched among the investigated $\alpha$-elements, and has the lowest dispersion. The Ti abundance derived from \ion{Ti}{i} lines is consistent with that of Mg and Ca, whereas the Ti abundance derived from \ion{Ti}{ii} is 0.1 dex lower but still within the statistical uncertainties. IP~Eri is slightly enriched in $\alpha$-elements ([$\alpha$/Fe]$ = 0.17\pm 0.06$ when considering Mg, Ca and Ti). The iron abundance determination is illustrated on Fig.~\ref{fig:fe_abu} and is consistent with the derived metallicity. 

\subsection{\textit{s}-process elements}
\label{Sect:s}
Abundances for elements from the first two \textit{s}-process peaks are measurable in IP~Eri and their values are listed in Table~\ref{tab:abu}. Examples of synthetic spectra are shown around two second-peak \textit{s}-process elements (see Figs.~\ref{fig:ba_spec} and~\ref{fig:ce_spec}) with solar ([Ba,Ce/Fe]= 0.00), enhanced ([Ba,Ce/Fe]= 1.00) and actual abundances ([Ba/Fe$]=0.18$ and [Ce/Fe$]=-0.22$). For elements from the first \textit{s}-process peak, lines of neutral and singly ionized Y and Zr are available. Abundances from neutral and ionized species agree within the statistical uncertainties excepted for Y, with abundances from neutral lines being somewhat larger than those derived from ionized lines. This trend, if real, could not be explained by the NLTE mechanism of overionization of the dominant neutral species due to UV radiation of non-local origin as shown, e.g., for Mg by \citet{Merle2011} or for Fe by \citet{Lind2012}. These NLTE effects altering the ionization equilibrium of \textit{s}-process elements  should be investigated, but such an analysis is beyond the scope of this paper.  

An average light-\textit{s}-process abundance  of [ls/Fe$]=0.06\pm0.04$ is obtained, based on \ion{Sr}{i}, \ion{Y}{ii}, and \ion{Zr}{ii} abundances, as compared to [hs/Fe$]=-0.03\pm0.05$,  based on \ion{Ba}{ii}, \ion{La}{ii}, and \ion{Ce}{ii} abundances. The subsolar  \ion{Ce}{ii} abundance is supported by a similarly subsolar   Nd abundance, as derived from \ion{Nd}{ii} lines. Thus, there does not seem to be a significant \textit{s}-process enrichment in IP~Eri. This issue is discussed further in Sect.~\ref{Sect:evolgal}, where the IP~Eri abundances are confronted with abundances in barium stars and with expectations from the chemical evolution of the Galaxy.

\begin{figure}
\begin{center}
 \includegraphics[width=0.9\linewidth]{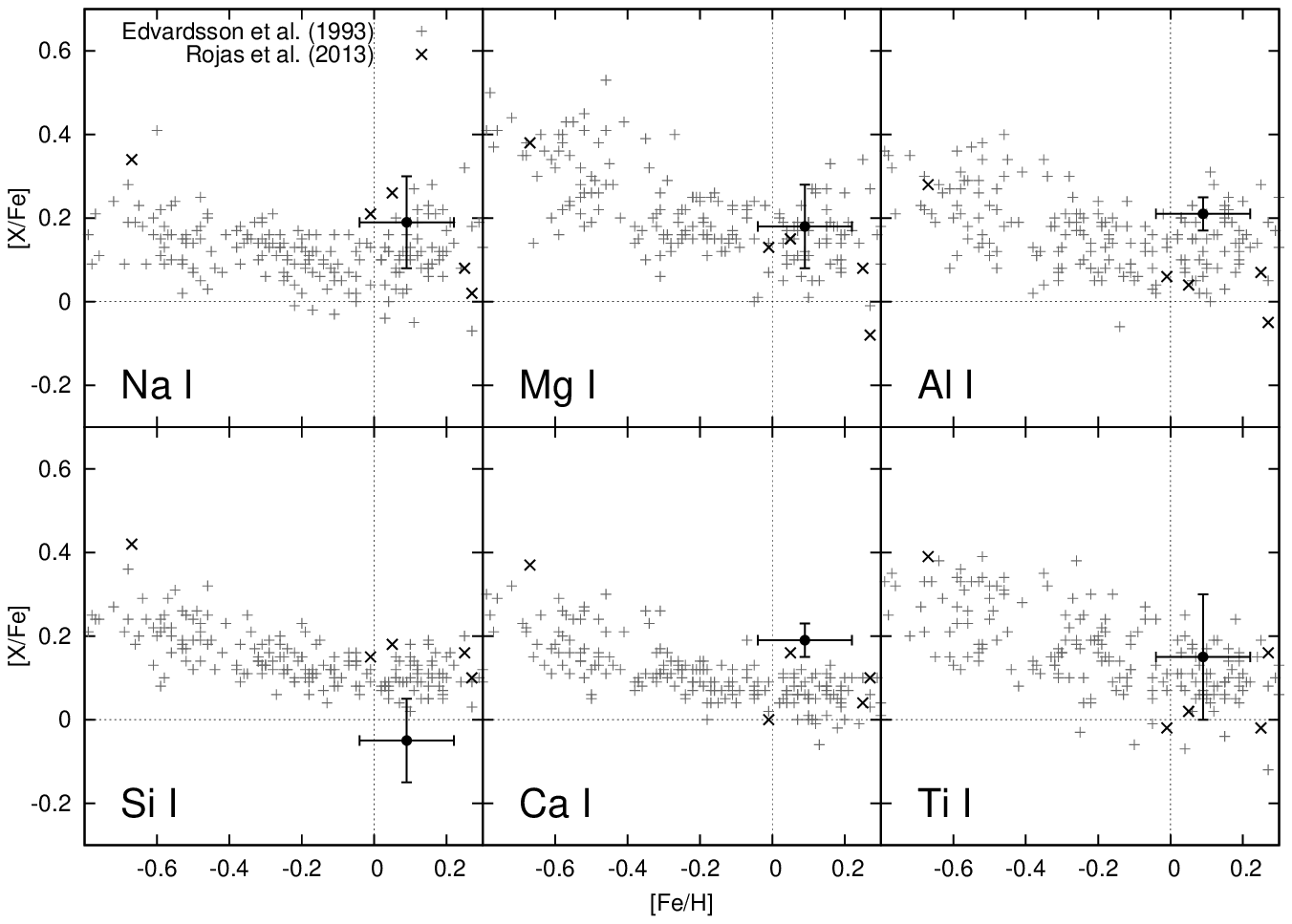}
 \vspace{0.2cm}
 \caption{Comparison of light-element abundances in IP Eri (black dot), in  field stars \citep[][grey plusses]{Edvardsson93}, and in mild barium stars \citep[][black crosses]{Rojas13}.}
 \label{fig:comp_light}
 \end{center}
\end{figure}

\section{Confrontation with normal field stars and barium stars}
\label{Sect:evolgal}

We have compared the abundances of IP~Eri with those of field stars and with those from a sample of mild barium stars, to confirm that the  abundances of \textit{s}-process elements in IP~Eri derived in Sect.~\ref{Sect:s} are by no means peculiar. 

We first consider the light-element abundances [X/Fe]  in IP~Eri which are compared in Fig.~\ref{fig:comp_light} with abundances in a large sample of  F and G stars from the galactic disc \citep{Edvardsson93}. We corrected for the zero-point solar abundance offset, since \citet{Edvardsson93} used a different reference value of $A_\odot(\mathrm{Fe})=7.51$. Fig.~\ref{fig:comp_light} shows that the light-element abundances in IP~Eri are in relatively good agreement with those of moderately metal-rich stars. The \ion{Na}{i}, \ion{Mg}{i}, \ion{Al}{i} and \ion{Ti}{i} abundances of IP~Eri are located within the abundance distribution of the galactic sample. \ion{O}{i} (not shown in Fig.~\ref{fig:comp_light}) and \ion{Ca}{i} are slightly more abundant as compared to the disc stars. The oxygen overabundance may be trusted though, since \citet{Edvardsson93} used a scaling relation to transform abundances from the high excitation lines that they used ($\lambda$615.8~nm as well as the $\lambda$777.3~nm triplet) to the abundance from the [\ion{O}{i}] $\lambda$630.0~nm line that we used. For Ca, there are no lines in common with the Edvardsson et al. analysis. We used saturated and strong lines which are affected by NLTE effects, as shown by \citet{Mashonkina07}. These authors predict positive NLTE abundance corrections (between 0 and 0.07~dex for a model with \Teff$=5000$~K, $\log{g}=3$ and \FeH$=0$), which, if accounted for, would further strengthen the Ca enrichment as compared to the \citet{Edvardsson93} values for field stars. We have also compared IP~Eri light-element abundances with those of a sample of mild barium stars from \citet{Rojas13}. Their light-element abundances are similar to those of field stars.

\begin{figure}
\begin{center}
 \includegraphics[width=0.9\linewidth]{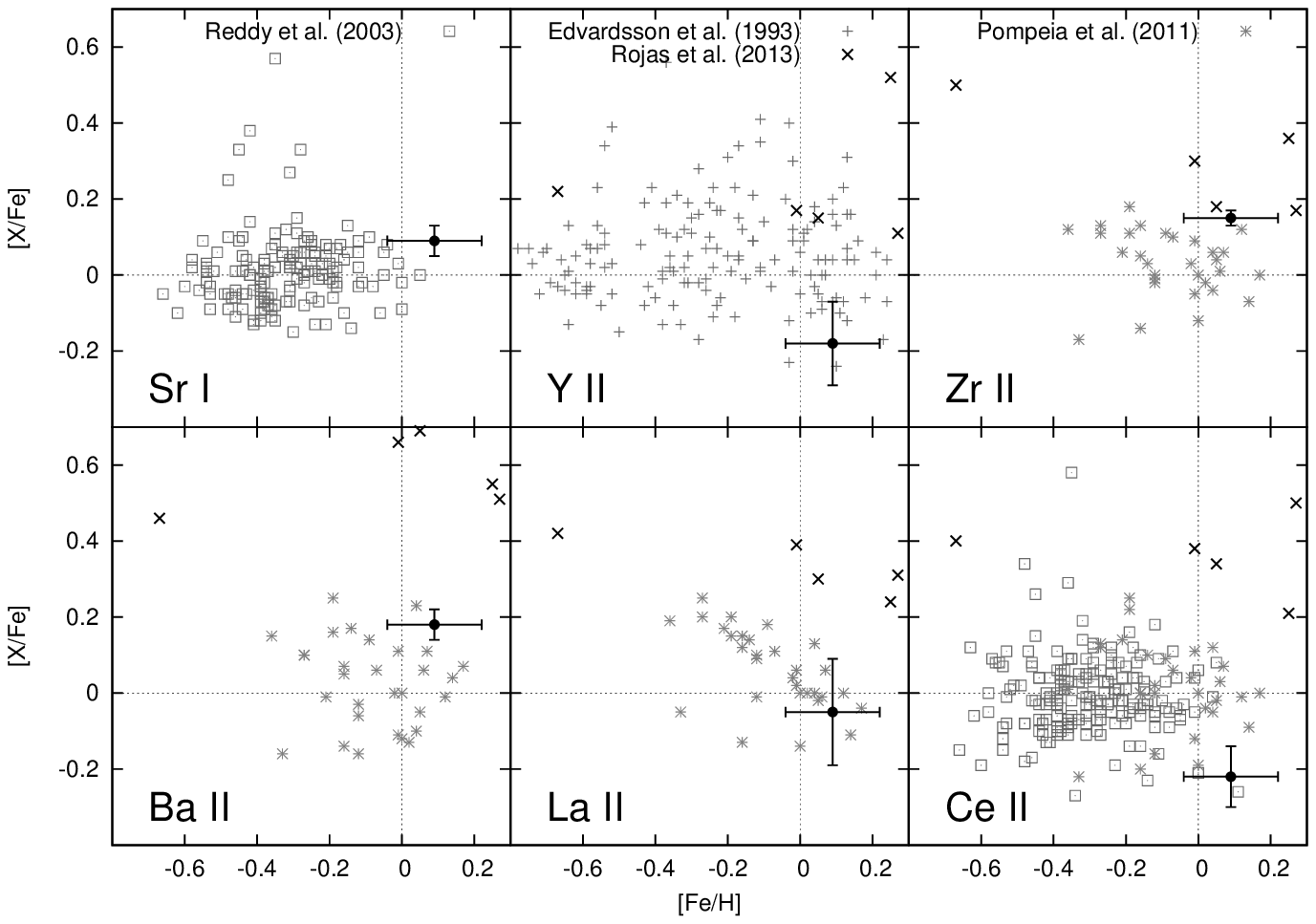}
  \vspace{0.2cm}
 \caption{Comparison of \textit{s}-process element abundances of IP Eri with field stars \citep[][grey plusses]{Edvardsson93} , \citep[][grey squares]{Reddy03},  \citep[][grey asterisks]{Pompeia11} and mild barium stars \citep[][black crosses]{Rojas13}.}
\label{fig:comp_s}
 \end{center}
\end{figure}

On the contrary, the \ion{Si}{i} abundance of IP~Eri (for which most of the lines used are common to the two studies) is lower than the value derived for the disc-star distribution. The IP~Eri abundance of \ion{Ti}{i} matches the average abundance of the disc stars of similar metallicities, but our value suffers from large error bars due to a larger set of lines in our analysis (9 against 4 for \citealt{Edvardsson93}). This large dispersion may be partially explained by the fact that all the lines have equivalent widths larger than 150 m\AA\ and are consequently less sensitive to abundance changes.

The \textit{s}-process element abundances in IP~Eri are compared with those of disc stars in Fig.~\ref{fig:comp_s}. The comparison sample is from \citet{Edvardsson93} for \ion{Y}{ii}, \citet{Reddy03} for \ion{Sr}{i} and \ion{Ce}{ii}, \citet{Pompeia11} for \ion{Zr}{ii}, \ion{Ba}{ii} and \ion{La}{ii}. The slight enrichment in \ion{Zr}{ii}, and \ion{Ba}{ii}  observed in IP~Eri is typical of the enrichment trend observed for disc stars. \ion{Y}{ii} and \ion{Ce}{ii} appear peculiar in that they are underabundant in IP~Eri as compared to disc stars of similar metallicities. 

For \ion{Sr}{i}, the only abundances available for comparison are from \citet{Reddy03}. They are based on a single line, and unfortunately, the Reddy et al. sample includes only stars with solar and sub-solar metallicities, giving the false impression that the IP~Eri abundances are discrepant. 
For \ion{La}{ii}, data from \citet{Pompeia11} are the only ones available for comparison, and the slight \ion{La}{ii} underabundance (with respect to the Sun) observed in IP~Eri matches the trend observed among disc stars. 

The comparison with the sample of mild barium stars of \citet{Rojas13} is not discriminating for first-peak \textit{s}-process elements (top panels of Fig.~\ref{fig:comp_s}). But abundances from second-peak \textit{s}-process elements in IP~Eri are clearly lower than those of mild barium stars (bottom panels of Fig.~\ref{fig:comp_s}). From these comparisons, we conclude that the slight enrichment in $\alpha$ and \textit{s}-process elements in IP~Eri is largely consistent with the chemical evolution of the Galaxy, so that there is no obvious signature from a chemical pollution resulting from mass transfer.   

\section{The evolutionary context of IP~Eri}
\label{Sect:analysis}

\begin{figure}[t]
\includegraphics[width=9cm]{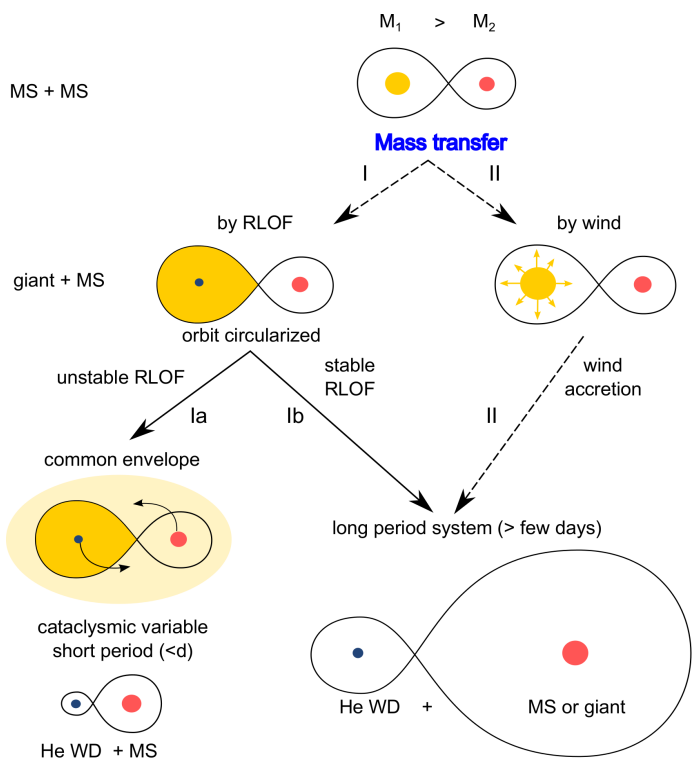}
\caption[]{\label{Fig:RLOF}
Evolutionary channels for the formation of a He WD. The dashed lines refer to channels where the eccentricity can be preserved (see text for details).}
\end{figure}

He WDs form when a star loses its hydrogen-rich envelope before it ignites helium. For a single star, this is not possible within a Hubble time since
only stars with masses $M\la 0.45-0.5$~M$_\odot$ can avoid helium ignition. A binary scenario is therefore required.

As illustrated in Fig.~\ref{Fig:RLOF}, different evolutionary channels can account for these objects. The first one involves mass transfer by Roche lobe overflow (RLOF). The long period of IP~Eri (1071 d) imposes that mass transfer starts while the star is already on the red-giant branch (late case B). Because of the presence of a deep convective envelope in the donor, two outcomes are possible depending on the mass ratio $q = M_{\rm donor}/M_{\rm gainer}$. If $q > 1.3-1.5$ \citep{Soberman1997,Hurley2002}, the mass transfer is dynamically unstable; after a rapid stage of common-envelope evolution, a short-period system forms (channel Ia in Fig.~\ref{Fig:RLOF}). In the alternative configuration ($q < 1.3-1.5$), soon after RLOF starts, the mass ratio reverses and subsequent mass transfer leads to the expansion of the orbit. The outcome is then a long-period system similar to IP~Eri (channel Ib). However, in this RLOF scenario, tidal interactions are very strong because of the extended convective envelope of the Roche-filling donor star and the orbit always circularizes. This channel thus cannot explain the high eccentricity of IP~Eri.

A solution to the eccentricity problem has been described in a companion paper \citep{Siess2014}. Based on binary-evolution calculations with the code BINSTAR \citep{Siess2013,Davis2013,Deschamps2013}, we showed that if the envelope of the He-WD progenitor is lost via tidally-enhanced winds (channel II), the circularization can be avoided. \citet{CRAP88} suggested that the presence of a companion star can substantially increase the mass-loss rate of the evolved component. In this situation, the donor star loses its envelope while remaining inside its Roche potential and tidal forces are significantly reduced. Moreover, if the system has an initial eccentricity, the orbital wind mass-transfer modulation \citep{Soker2000} provides an eccentricity-pumping mechanism that counteracts the tidal circularization. We showed that such a scenario is able to account for all the orbital properties of a system like IP~Eri.

\begin{figure}[t]
\includegraphics[width=9cm]{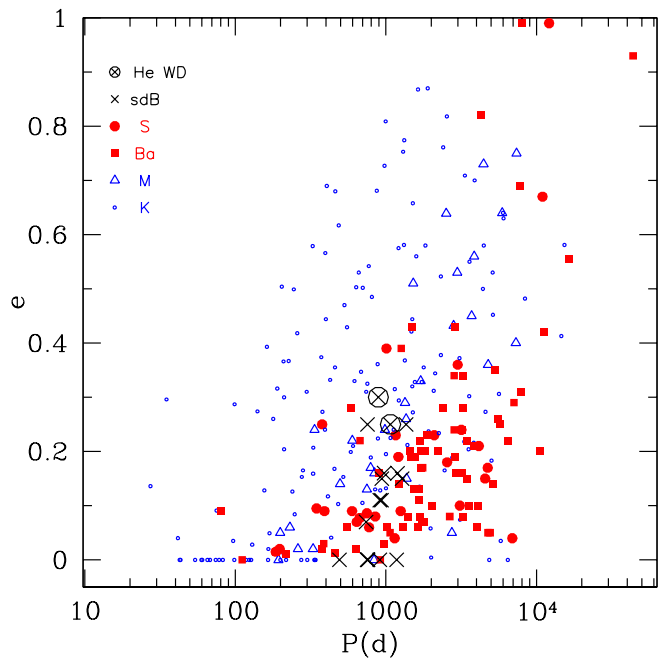}
\caption[]{\label{Fig:eP}
The positions of IP~Eri and its twin system HR~1608 \citep[black circled  crosses; Beavers \& Eitter 1988, quoted by][]{Landsman1993} in the eccentricity -- period diagram are compared with those of the long-period sdB binaries, with (mostly) pre-mass-transfer binaries (normal K and M giants), and with post-mass-transfer binaries (Barium and S stars). Symbols are as indicated in the figure label.}
\end{figure}

In Fig.~\ref{Fig:eP} we compare the orbital properties of IP~Eri with those of pre- \citep[K giants in open clusters from Mermilliod et al. 2007 and binary M giants from][]{Famaey2009,Jorissen2009} \nocite{Mermilliod2007} and post-mass-transfer binaries \citep[barium and S stars from][updated with some recently published orbits from \citealt{Gorlova2014}]{Jorissen1998}. We also show in the $e-\log P$ diagram the location of the G5 IV + WD system HR~1608 (= 63~Eri). This system has an eccentricity ($e=0.30\pm0.06$) and a period ($P=903\pm5$~d) very similar to those of IP~Eri \citep[from][ quoted by Landsman et al. 1993]{Beavers1988}\nocite{Landsman1993}. The revised Hipparcos parallax \citep[$18.53\pm0.84$~mas;][]{vanLeeuwen2007} yields a distance range 51 -- 56~pc for HR~1608, which implies a WD mass around 0.4~M$_{\odot}$ \citep{Landsman1993}, and thus a He-WD. A more recent analysis of the WD parameters by \citet{Vennes1998} suggests instead a higher mass range (0.51 -- 0.67~M$_{\odot}$). We also include in Fig.~\ref{Fig:eP} the sdB binaries with long periods (black crosses) from \citet{Ostensen2011,Ostensen2012}, \citet{Vos2012,Vos2013}, \citet{Deca2012}, and \citet{Barlow2012,Barlow2013}.

IP~Eri and HR~1608 have the largest eccentricities when compared to the long-period sdB systems, a fact whose significance is difficult to assess with so few He-WD systems.  What seems significant, however, is the fact that (long-period) sdB and He-WD systems occupy a rather restricted period range around $10^3$~d. In that respect, they differ from the barium and S binaries, two other families of post-mass-transfer systems, which spread over a more extended period range. This difference might be related to the fact that sdB and He-WD binaries are the end products of mass transfer occurring on the first red giant branch, whereas barium and S systems involved mass transfer on the AGB. 

\section{Summary}
\label{Sect:summary}

IP~Eri is an unusual long-period binary system with a high eccentricity. Our abundance analysis reveals that it is not enriched in \textit{s}-process elements, giving additional support to He-WD nature of the hot component that avoided evolution along the AGB. The abundances of the light elements do not reveal any peculiarities and within the error bars, the abundances are very close to solar. What is more surprising is that IP~Eri alike its twin HR~1608 have the largest eccentricities among their closely related sdB systems and are very concentrated around a thousand-day period, which is likely related to the fact that the donor stars lose their envelope on the red-giant branch. However, the statistics remains small and additional data are highly desirable in order to confirm the formation channel of these systems.

\begin{acknowledgement}
This work has been partly funded by an {\it Action de recherche concertée} (ARC) from the {\it Direction g\'en\'erale de l'Enseignement non obligatoire et de la Recherche scientifique -- Direction de la recherche scientifique -- Communaut\'e fran\c{c}aise de Belgique.} T.M. is supported by the FNRS-F.R.S. as temporary post-doctoral researcher  under grant No. 2.4513.11. The Mercator telescope is operated thanks to grant number G.0C31.13 of the FWO under the ’Big Science’ initiative of the Flemish governement. Based on observations obtained with the HERMES spectrograph, supported by the Fund for Scientific Research of Flanders (FWO), the Research Council of K.U.Leuven, the Fonds National de la Recherche Scientifique (F.R.S.-FNRS), Belgium, the Royal Observatory of Belgium, the Observatoire de Gen\`eve, Switzerland and the Th\"uringer Landessternwarte Tautenburg, Germany.
\end{acknowledgement}

\bibliographystyle{aa}
\bibliography{biblio}

\onecolumn
\appendix
\section{Linelist}
\label{ap:ll}
\tiny
\begin{longtab}
 \begin{longtable}{lcrcc}
\caption{Line list used for determining atmospheric parameters and chemical composition of  
atomic species. The main source for oscillator strengths ($\log{gf}$) is VALD. 
Fitted individual absolute abundances $A(X)$ are also given. 
The absolute solar abundances $A_\odot(X)$ are from \citet{Grevesse07}. 
Vertical lines in the left margin mean that all transitions contribute to the same line 
(fine or hyperfine structure). For \ion{Ba}{ii}, the atomic mass of the contributing isotope 
is indicated as a superscript to the transition wavelength.}\\
\tiny
 $\lambda$ [nm] & $\chi$ [eV] & $\log{gf}$ & $A(X)$ \\
 \hline
 \endfirsthead
 \caption{Continued.}\\
 $\lambda$ [nm] & $\chi$ [eV] & $\log{gf}$ & $A(X)$ \\
 \hline
 \endhead
 \hline
 \endfoot
\\
\textbf{C I} & \multicolumn{3}{l}{A$_\odot$(C) = 8.39}\\
493.2049 &  7.685 & $-1.884$ & 8.25 \\ 
538.0337 &  7.685 & $-1.615$ & 8.63 \\ 
555.3174 &  8.643 & $-2.370$ & 8.45 \\ 
633.5701 &  8.771 & $-2.370$ & 8.63 \\ 
633.7183 &  8.771 & $-2.450$ & 8.65 \\ 
658.7610 &  8.537 & $-1.021$ & 8.62 \\ 
661.1353 &  8.851 & $-1.837$ & 8.64 \\ 
667.1845 &  8.851 & $-1.651$ & 8.61 \\ 
768.5190 &  8.771 & $-1.519$ & 8.58 \\ 
784.8241 &  8.848 & $-1.731$ & 8.68 \\ 
785.2859 &  8.851 & $-1.683$ & 8.74 \\ 
786.0877 &  8.851 & $-1.148$ & 8.65 \\ 
788.4490 &  8.847 & $-1.580$ & 8.31 \\ 
807.8479 &  8.848 & $-1.817$ & 8.69 \\ 
833.5148 &  7.685 & $-0.420$ & 8.60 \\ 
872.7126 &  1.264 & $-8.136$ & 8.61 \\ 
881.8479 &  9.003 & $-1.060$ & 8.20 \\
\\
\textbf{O I} & \multicolumn{3}{l}{A$_\odot$(O) = 8.66}\\
630.0304 &  0.000 & $-9.715$ & 8.82 \\
\\
\textbf{Na I} & \multicolumn{3}{l}{A$_\odot$(Na) = 6.17}\\
449.7657 & 2.104 & $-1.560$ & 6.36 \\
498.2814 & 2.104 & $-0.950$ & 6.38 \\
568.2633 & 2.102 & $-0.706$ & 6.25 \\
568.8205 & 2.104 & $-0.450$ & 6.45 \\
615.4226 & 2.102 & $-1.547$ & 6.53 \\
616.0747 & 2.104 & $-1.246$ & 6.48 \\
818.3255 & 2.102 & $ 0.230$ & 6.60 \\
819.4824 & 2.104 & $ 0.490$ & 6.54 \\
\\
\textbf{Mg I} & \multicolumn{3}{l}{A$_\odot$(Mg) = 7.53}\\
552.8405 & 4.346 & $-0.620$ & 7.70 \\
571.1088 & 4.346 & $-1.833$ & 7.67 \\
631.8717 & 5.108 & $-2.103$ & 7.86 \\
631.9237 & 5.108 & $-2.324$ & 7.92 \\
631.9495 & 5.108 & $-2.803$ & 7.89 \\
738.7689 & 5.753 & $-1.100$ & 7.82 \\
769.1553 & 5.753 & $-0.783$ & 7.68 \\
\\
\vline~871.2676 & 5.932 & $-1.670$ & 7.80 \\
\vline~871.2689 & 5.932 & $-1.370$ \\
\\   
871.7825 & 5.933 & $-0.930$ & 7.75 \\
\\
\vline~873.6019 & 5.946 & $-0.690$ & 7.96 \\ 
\vline~873.6029 & 5.946 & $-1.020$ & \\
\\
\textbf{Al I} & \multicolumn{3}{l}{A$_\odot$(Al) = 6.37}\\
555.7063 & 6.143 & $-2.104$ & 6.63 \\
669.6023 & 3.143 & $-1.569$ & 6.69 \\
669.8673 & 3.143 & $-1.870$ & 6.73 \\
783.5309 & 4.022 & $-0.649$ & 6.68 \\
783.6134 & 4.022 & $-0.494$ & 6.71 \\
877.2865 & 4.022 & $-0.170$ & 6.64 \\
877.3896 & 4.022 & $-0.161$ & 6.75 \\
\\
\textbf{Si I} & \multicolumn{3}{l}{A$_\odot$(Si) = 7.51}\\
566.5555 & 4.920 & $-1.940$ & 7.31 \\
568.4484 & 4.954 & $-1.553$ & 7.55 \\
569.0425 & 4.930 & $-1.773$ & 7.50 \\
570.1104 & 4.930 & $-1.953$ & 7.57 \\
577.2146 & 5.082 & $-1.653$ & 7.64 \\
579.3073 & 4.930 & $-1.963$ & 7.62 \\
612.5021 & 5.614 & $-1.464$ & 7.60 \\
613.1573 & 5.616 & $-1.556$ & 7.52 \\
613.1852 & 5.616 & $-1.615$ & 7.57 \\
614.2483 & 5.619 & $-1.295$ & 7.43 \\
614.5016 & 5.616 & $-1.310$ & 7.54 \\
615.5134 & 5.619 & $-0.754$ & 7.70 \\
624.4466 & 5.616 & $-1.093$ & 7.42 \\
776.0628 & 6.206 & $-1.261$ & 7.65 \\
872.8010 & 6.181 & $-0.370$ & 7.61 \\
\\
\textbf{Ca I} & \multicolumn{3}{l}{A$_\odot$(Ca) = 6.31}\\
558.8749 & 2.526 & $ 0.358$ & 6.62 \\
585.7451 & 2.933 & $ 0.240$ & 6.50 \\
610.2723 & 1.879 & $-0.793$ & 6.60 \\
612.2217 & 1.886 & $-0.316$ & 6.59 \\
616.2173 & 1.899 & $-0.090$ & 6.61 \\
643.9075 & 2.526 & $ 0.390$ & 6.55 \\
649.3781 & 2.521 & $-0.109$ & 6.57 \\
649.9650 & 2.523 & $-0.818$ & 6.64 \\
\\
\textbf{Ti I} & \multicolumn{3}{l}{A$_\odot$(Ti) = 4.90}\\
453.4776 & 0.836 & $ 0.280$ & 5.31 \\
454.8763 & 0.826 & $-0.354$ & 4.99 \\
498.1731 & 0.848 & $ 0.560$ & 5.18 \\
499.9503 & 0.826 & $ 0.306$ & 5.24 \\
502.4844 & 0.818 & $-0.546$ & 5.05 \\ 
503.9957 & 0.021 & $-1.074$ & 5.02 \\  
517.3743 & 0.000 & $-1.062$ & 4.97 \\ 
519.2969 & 0.021 & $-1.006$ & 5.39 \\
521.0385 & 0.048 & $-0.527$ & 5.09 \\
\\
\textbf{Ti II} \\
453.3960 & 1.237 & $-0.530$ & 4.86 \\
456.3757 & 1.221 & $-0.690$ & 5.05 \\
457.1968 & 1.572 & $-0.320$ & 4.99 \\
533.6771 & 1.582 & $-1.630$ & 4.98 \\
538.1015 & 1.566 & $-1.970$ & 5.26 \\
\\
\textbf{Fe I} & \multicolumn{3}{l}{A$_\odot$(Fe) = 7.45}\\

480.8148 & 3.251  & $-2.690$ & 7.58 \\ 
496.2572 & 4.178  & $-1.182$ & 7.48 \\ 
499.2785 & 4.260  & $-2.350$ & 7.71 \\ 
505.8496 & 3.642  & $-2.830$ & 7.73 \\ 
524.3776 & 4.256  & $-1.050$ & 7.51 \\ 
525.3021 & 2.279  & $-3.840$ & 7.50 \\ 
529.4547 & 3.640  & $-2.760$ & 7.62 \\ 
529.5312 & 4.415  & $-1.590$ & 7.55 \\ 
537.3709 & 4.473  & $-0.760$ & 7.45 \\ 
537.9574 & 3.694  & $-1.514$ & 7.48 \\ 
538.9479 & 4.415  & $-0.410$ & 7.24 \\ 
539.7618 & 3.634  & $-2.528$ & 7.68 \\ 
539.8279 & 4.445  & $-0.630$ & 7.65 \\ 
541.2784 & 4.434  & $-1.716$ & 7.41 \\ 
541.7033 & 4.415  & $-1.580$ & 7.43 \\ 
543.6295 & 4.386  & $-1.440$ & 7.65 \\ 
544.1339 & 4.312  & $-1.630$ & 7.36 \\ 
546.6396 & 4.371  & $-0.630$ & 7.62 \\ 
547.3163 & 4.191  & $-2.040$ & 7.53 \\ 
548.3099 & 4.154  & $-1.406$ & 7.57 \\ 
548.7145 & 4.415  & $-1.430$ & 7.43 \\ 
549.1832 & 4.186  & $-2.188$ & 7.51 \\ 
549.4463 & 4.076  & $-1.990$ & 7.54 \\ 
552.2446 & 4.209  & $-1.450$ & 7.47 \\ 
553.9280 & 3.642  & $-2.560$ & 7.22 \\ 
554.3936 & 4.217  & $-1.040$ & 7.36 \\ 
554.9949 & 3.694  & $-2.810$ & 7.50 \\ 
556.0212 & 4.434  & $-1.090$ & 7.50 \\ 
557.7025 & 5.033  & $-1.543$ & 7.69 \\ 
561.8632 & 4.209  & $-1.275$ & 7.46 \\ 
563.3946 & 4.991  & $-0.230$ & 7.38 \\ 
563.8262 & 4.220  & $-0.770$ & 7.63 \\ 
565.1469 & 4.473  & $-1.900$ & 7.75 \\ 
565.2318 & 4.260  & $-1.850$ & 7.68 \\ 
566.1345 & 4.284  & $-1.756$ & 7.52 \\ 
567.9023 & 4.652  & $-0.820$ & 7.53 \\ 
570.5464 & 4.301  & $-1.355$ & 7.46 \\ 
573.1762 & 4.256  & $-1.200$ & 7.66 \\ 
573.2296 & 4.991  & $-1.460$ & 7.67 \\ 
574.1848 & 4.256  & $-1.672$ & 7.57 \\ 
575.2032 & 4.549  & $-1.177$ & 7.78 \\ 
577.5081 & 4.220  & $-1.297$ & 7.75 \\ 
577.8453 & 2.588  & $-3.430$ & 7.46 \\ 
584.9683 & 3.694  & $-2.890$ & 7.57 \\ 
585.5076 & 4.608  & $-1.478$ & 7.49 \\ 
585.8778 & 4.220  & $-2.160$ & 7.49 \\ 
586.1109 & 4.283  & $-2.304$ & 7.44 \\ 
590.5671 & 4.652  & $-0.690$ & 7.25 \\ 
592.7789 & 4.652  & $-0.990$ & 7.41 \\ 
593.0180 & 4.652  & $-0.230$ & 7.70 \\ 
593.4655 & 3.928  & $-1.070$ & 7.50 \\ 
595.6694 & 0.859  & $-4.553$ & 7.54 \\ 
602.7051 & 4.076  & $-1.089$ & 7.46 \\ 
605.6005 & 4.733  & $-0.460$ & 7.58 \\ 
609.3643 & 4.607  & $-1.400$ & 7.62 \\ 
615.1617 & 2.176  & $-3.312$ & 7.54 \\ 
616.5360 & 4.143  & $-1.473$ & 7.48 \\ 
617.3334 & 2.223  & $-2.880$ & 7.64 \\ 
618.7989 & 3.943  & $-1.620$ & 7.50 \\ 
620.0313 & 2.608  & $-2.405$ & 7.72 \\ 
622.6734 & 3.883  & $-2.120$ & 7.59 \\ 
627.0223 & 2.858  & $-2.536$ & 7.43 \\ 
632.2685 & 2.588  & $-2.448$ & 7.74 \\ 
643.6406 & 4.186  & $-2.580$ & 7.80 \\ 
647.5624 & 2.559  & $-2.941$ & 7.79 \\ 
648.1870 & 2.279  & $-2.985$ & 7.57 \\ 
649.8938 & 0.958  & $-4.688$ & 7.48 \\ 
651.8366 & 2.831  & $-2.373$ & 7.36 \\ 
665.3851 & 4.154  & $-2.215$ & 7.42 \\ 
669.9141 & 4.593  & $-2.101$ & 7.53 \\ 
671.0318 & 1.485  & $-4.764$ & 7.48 \\ 
671.3743 & 4.795  & $-1.500$ & 7.69 \\ 
672.5356 & 4.103  & $-2.013$ & 7.42 \\ 
672.6666 & 4.607  & $-1.133$ & 7.67 \\ 
673.9521 & 1.557  & $-4.794$ & 7.46 \\ 
675.0152 & 2.424  & $-2.604$ & 7.69 \\ 
681.0262 & 4.607  & $-0.986$ & 7.55 \\ 
848.1981 & 4.186  & $-1.988$ & 7.51 \\ 
851.5108 & 3.018  & $-2.073$ & 7.72 \\ 
852.7852 & 5.020  & $-1.625$ & 7.73 \\ 
857.1804 & 5.010  & $-1.414$ & 7.91 \\ 
859.8829 & 4.386  & $-1.088$ & 7.49 \\ 
862.1601 & 2.949  & $-2.320$ & 7.67 \\ 
863.2414 & 4.103  & $-2.341$ & 7.53 \\ 
869.8706 & 2.990  & $-3.452$ & 7.74 \\ 
869.9454 & 4.955  & $-0.380$ & 7.51 \\ 
880.4623 & 2.279  & $-3.234$ & 7.66 \\ 
\\
\textbf{Fe II}\\
499.3358 & 2.807  & $-3.684$ & 7.32 \\
523.4625 & 3.221  & $-2.180$ & 7.60 \\
525.6938 & 2.891  & $-4.182$ & 7.64 \\
532.5553 & 3.221  & $-3.160$ & 7.44 \\
533.7732 & 3.230  & $-3.720$ & 7.63 \\
541.4073 & 3.221  & $-3.580$ & 7.35 \\
542.5257 & 3.199  & $-3.220$ & 7.29 \\
553.4847 & 3.245  & $-2.865$ & 7.48 \\
599.1376 & 3.153  & $-3.647$ & 7.63 \\
608.4111 & 3.199  & $-3.881$ & 7.59 \\
611.3322 & 3.221  & $-4.230$ & 7.59 \\
614.9258 & 3.889  & $-2.841$ & 7.66 \\
624.7557 & 3.892  & $-2.435$ & 7.52 \\
636.9462 & 2.891  & $-4.110$ & 7.44 \\
643.2680 & 2.891  & $-3.570$ & 7.54 \\
645.6383 & 3.903  & $-2.185$ & 7.71 \\
\\
\textbf{Sr I} & \multicolumn{3}{l}{A$_\odot$(Sr) = 2.92}\\
483.2108 & 1.798 & $-0.110$ & 3.11 \\ 
496.2259 & 1.847 & $ 0.200$ & 3.06 \\
707.0070 & 1.847 & $-0.030$ & 3.13 \\
\\
\textbf{Y I} & \multicolumn{3}{l}{A$_\odot$(Y) = 2.21}\\
619.1718 & 0.000 & $-0.970$ & 2.38 \\ 
643.5004 & 0.066 & $-0.820$ & 2.28 \\
\\
\textbf{Y II} \\
490.0120 & 1.033 & $ 0.103$ & 1.84 \\ 
520.0406 & 0.992 & $-0.570$ & 2.01 \\ 
520.5724 & 1.033 & $-0.193$ & 2.00 \\
528.9815 & 1.033 & $-1.850$ & 2.03 \\
679.5414 & 1.738 & $-1.030$ & 1.80 \\
\\
\textbf{Zr I} & \multicolumn{3}{l}{A$_\odot$(Zr) = 2.58}\\
482.8041 & 0.623 & $-0.640$ & 2.81 \\ 
612.7475 & 0.154 & $-1.060$ & 2.82 \\ 
613.4585 & 0.000 & $-1.280$ & 2.76 \\
807.0115 & 0.730 & $-0.790$ & 3.16 \\
\\
\textbf{Zr II} \\
444.2992 & 1.486 & $-0.420$ & 2.82 \\
535.0089 & 1.827 & $-1.240$ & 2.81 \\
\\
\textbf{Ba II} & \multicolumn{3}{l}{A$_\odot$(Ba) = 2.17}\\
\vline~455.3998$^{137}$ & 0.000 & -0.666 & \\ 
\vline~455.3999$^{137}$ & 0.000 & -0.666 & \\ 
\vline~455.4000$^{137}$ & 0.000 & -1.064 & \\ 
\vline~455.4001$^{135}$ & 0.000 & -0.666 & \\ 
\vline~455.4002$^{135}$ & 0.000 & -1.064 & \\ 
\vline~455.4002$^{135}$ & 0.000 & -0.666 & \\ 
\vline~455.4031$^{130}$ & 0.000 &  0.140 & \\ 
\vline~455.4031$^{132}$ & 0.000 &  0.140 & \\ 
\vline~455.4031$^{134}$ & 0.000 &  0.140 & 2.42\\ 
\vline~455.4032$^{136}$ & 0.000 &  0.140 & \\ 
\vline~455.4033$^{138}$ & 0.000 &  0.140 & \\ 
\vline~455.4048$^{135}$ & 0.000 & -0.219 & \\ 
\vline~455.4050$^{135}$ & 0.000 & -0.666 & \\ 
\vline~455.4051$^{137}$ & 0.000 & -0.219 & \\ 
\vline~455.4052$^{135}$ & 0.000 & -1.365 & \\ 
\vline~455.4054$^{137}$ & 0.000 & -0.666 & \\ 
\vline~455.4055$^{137}$ & 0.000 & -1.365 & \\ 
\\
\vline~585.3669$^{135}$ & 0.604 & $-1.967$ & \\ 
\vline~585.3669$^{137}$ & 0.604 & $-1.967$ & \\ 
\vline~585.3670$^{135}$ & 0.604 & $-2.113$ & \\ 
\vline~585.3670$^{135}$ & 0.604 & $-1.909$ & \\ 
\vline~585.3671$^{137}$ & 0.604 & $-2.113$ & \\ 
\vline~585.3671$^{137}$ & 0.604 & $-1.909$ & \\ 
\vline~585.3672$^{135}$ & 0.604 & $-2.113$ & \\ 
\vline~585.3672$^{135}$ & 0.604 & $-2.511$ & \\ 
\vline~585.3673$^{130}$ & 0.604 & $-0.909$ & \\ 
\vline~585.3673$^{132}$ & 0.604 & $-0.909$ & \\ 
\vline~585.3673$^{134}$ & 0.604 & $-0.909$ & \\ 
\vline~585.3673$^{137}$ & 0.604 & $-2.113$ & 2.43 \\ 
\vline~585.3673$^{135}$ & 0.604 & $-1.812$ & \\ 
\vline~585.3673$^{137}$ & 0.604 & $-2.511$ & \\ 
\vline~585.3674$^{136}$ & 0.604 & $-0.909$ & \\
\vline~585.3675$^{135}$ & 0.604 & $-1.909$ & \\ 
\vline~585.3675$^{135}$ & 0.604 & $-1.365$ & \\
\vline~585.3675$^{137}$ & 0.604 & $-1.812$ & \\ 
\vline~585.3675$^{138}$ & 0.604 & $-0.909$ & \\
\vline~585.3676$^{137}$ & 0.604 & $-1.909$ & \\ 
\vline~585.3676$^{137}$ & 0.604 & $-1.365$ & \\
\vline~585.3680$^{135}$ & 0.604 & $-1.967$ & \\ 
\vline~585.3682$^{137}$ & 0.604 & $-1.967$ & \\ 
\\
\vline~614.1708$^{135}$ & 0.704 & $-0.456$ &  \\ 
\vline~614.1708$^{135}$ & 0.704 & $-1.264$ &  \\  
\vline~614.1709$^{135}$ & 0.704 & $-2.410$ &  \\  
\vline~614.1709$^{137}$ & 0.704 & $-1.264$ &  \\  
\vline~614.1709$^{137}$ & 0.704 & $-0.456$ &  \\  
\vline~614.1710$^{137}$ & 0.704 & $-2.410$ &  \\  
\vline~614.1711$^{130}$ & 0.704 & $-0.030$ &  \\ 
\vline~614.1711$^{132}$ & 0.704 & $-0.030$ &  \\ 
\vline~614.1711$^{134}$ & 0.704 & $-0.030$ &  \\ 
\vline~614.1712$^{136}$ & 0.704 & $-0.030$ &  \\  
\vline~614.1713$^{135}$ & 0.704 & $-0.662$ &  \\  
\vline~614.1713$^{138}$ & 0.704 & $-0.030$ &  2.40 \\ 
\vline~614.1714$^{135}$ & 0.704 & $-1.167$ &  \\  
\vline~614.1715$^{135}$ & 0.704 & $-2.234$ &  \\  
\vline~614.1715$^{137}$ & 0.704 & $-0.662$ &  \\  
\vline~614.1716$^{135}$ & 0.704 & $-0.912$ &  \\  
\vline~614.1716$^{137}$ & 0.704 & $-1.167$ &  \\  
\vline~614.1717$^{135}$ & 0.704 & $-1.234$ &  \\  
\vline~614.1717$^{135}$ & 0.704 & $-1.280$ &  \\  
\vline~614.1717$^{137}$ & 0.704 & $-2.234$ &  \\  
\vline~614.1718$^{137}$ & 0.704 & $-0.912$ &  \\  
\vline~614.1719$^{137}$ & 0.704 & $-1.234$ &  \\  
\vline~614.1719$^{137}$ & 0.704 & $-1.280$ &  \\  
\\
\vline~6496.883$^{135}$ & 0.604 & $-1.911$ & \\ 
\vline~6496.883$^{137}$ & 0.604 & $-1.911$ & \\ 
\vline~6496.888$^{135}$ & 0.604 & $-1.212$ & \\ 
\vline~6496.888$^{137}$ & 0.604 & $-1.212$ & \\ 
\vline~6496.895$^{130}$ & 0.604 & $-0.406$ & \\
\vline~6496.895$^{132}$ & 0.604 & $-0.406$ & \\
\vline~6496.895$^{134}$ & 0.604 & $-0.406$ & \\
\vline~6496.895$^{135}$ & 0.604 & $-0.765$ & \\
\vline~6496.896$^{137}$ & 0.604 & $-0.765$ & 2.49\\
\vline~6496.897$^{136}$ & 0.604 & $-0.406$ & \\
\vline~6496.898$^{138}$ & 0.604 & $-0.406$ & \\
\vline~6496.900$^{135}$ & 0.604 & $-1.610$ & \\ 
\vline~6496.902$^{135}$ & 0.604 & $-1.212$ & \\ 
\vline~6496.902$^{137}$ & 0.604 & $-1.610$ & \\ 
\vline~6496.904$^{137}$ & 0.604 & $-1.212$ & \\ 
\vline~6496.906$^{135}$ & 0.604 & $-1.212$ & \\ 
\vline~6496.909$^{137}$ & 0.604 & $-1.212$ & \\
\\
\textbf{La II} & \multicolumn{3}{l}{A$_\odot$(La) = 1.13}\\
\vline~408.6695  & 0.000 &$-1.266$ & \\     
\vline~408.6699  & 0.000 &$-1.108$ & \\ 
\vline~408.6702  & 0.000 &$-1.119$ & \\    
\vline~408.6705  & 0.000 &$-1.292$ & \\    
\vline~408.6708  & 0.000 &$-0.696$ & \\     
\vline~408.6709  & 0.000 &$-1.094$ & \\     
\vline~408.6710  & 0.000 &$-1.790$ & 1.04 \\    
\vline~408.6711  & 0.000 &$-1.468$ & \\    
\vline~408.6711  & 0.000 &$-3.216$ & \\    
\vline~408.6717  & 0.000 &$-1.292$ & \\    
\vline~408.6719  & 0.000 &$-1.119$ & \\    
\vline~408.6720  & 0.000 &$-1.108$ & \\    
\vline~408.6721  & 0.000 &$-1.266$ & \\
\\
455.8457    & 0.321 &$-0.970$ & 1.15 \\  
457.4860    & 0.173 &$-1.08$  & 1.24 \\
\\
\vline~492.0965 & 0.126 & $-2.261$ & \\  
\vline~492.0965 & 0.126 & $-2.407$ & \\  
\vline~492.0966 & 0.126 & $-2.065$ & \\  
\vline~492.0966 & 0.126 & $-2.078$ & \\  
\vline~492.0966 & 0.126 & $-2.738$ & \\  
\vline~492.0968 & 0.126 & $-1.831$ & \\  
\vline~492.0968 & 0.126 & $-1.956$ & \\  
\vline~492.0968 & 0.126 & $-2.629$ & \\  
\vline~492.0971 & 0.126 & $-1.646$ & \\   
\vline~492.0971 & 0.126 & $-1.895$ &  1.32\\  
\vline~492.0971 & 0.126 & $-2.650$ & \\  
\vline~492.0975 & 0.126 & $-1.490$ & \\   
\vline~492.0975 & 0.126 & $-1.891$ & \\   
\vline~492.0975 & 0.126 & $-2.760$ & \\  
\vline~492.0979 & 0.126 & $-1.354$ & \\   
\vline~492.0979 & 0.126 & $-1.957$ & \\   
\vline~492.0979 & 0.126 & $-2.972$ & \\   
\vline~492.0985 & 0.126 & $-1.233$ & \\   
\vline~492.0985 & 0.126 & $-2.162$ & \\   
\vline~492.0985 & 0.126 & $-3.375$ & \\
\\
\vline~530.3513 & 0.321 & $-1.874$ & \\   
\vline~530.3513 & 0.321 & $-2.363$ & \\   
\vline~530.3514 & 0.321 & $-3.062$ & \\   
\vline~530.3531 & 0.321 & $-2.167$ & \\  
\vline~530.3532 & 0.321 & $-2.247$ & 1.32 \\  
\vline~530.3532 & 0.321 & $-2.622$ & \\  
\vline~530.3546 & 0.321 & $-2.366$ & \\  
\vline~530.3546 & 0.321 & $-2.622$ & \\  
\vline~530.3547 & 0.321 & $-2.351$ & \\ 
\\
579.7565 & 0.244 & $-1.360$ & 1.33 \\
588.0633 & 0.235 & $-1.830$ & 1.11 \\
632.0376 & 0.173 & $-1.520$ & 1.04\\ 
\\
\vline~639.0455 & 0.321 & $-2.012$ & \\ 
\vline~639.0468 & 0.321 & $-2.183$ & \\ 
\vline~639.0468 & 0.321 & $-2.752$ & \\ 
\vline~639.0479 & 0.321 & $-2.570$ & \\ 
\vline~639.0479 & 0.321 & $-3.752$ & \\ 
\vline~639.0480 & 0.321 & $-2.390$ & \\ 
\vline~639.0489 & 0.321 & $-2.536$ & \\
\vline~639.0489 & 0.321 & $-3.334$ & 1.23 \\
\vline~639.0490 & 0.321 & $-2.661$ & \\
\vline~639.0496 & 0.321 & $-3.100$ & \\
\vline~639.0497 & 0.321 & $-2.595$ & \\
\vline~639.0498 & 0.321 & $-3.079$ & \\
\vline~639.0502 & 0.321 & $-2.954$ & \\
\vline~639.0503 & 0.321 & $-2.778$ & \\
\vline~639.0506 & 0.321 & $-2.857$ & \\
\\
677.4268 & 0.126 & $-1.708$ & 0.92\\
\\
\textbf{Ce II} & \multicolumn{3}{l}{A$_\odot$(Ce) = 1.70}\\
407.3374 & 0.478 & $0.230$ & 1.52 \\
452.3075 & 0.516  & $-0.030$ & 1.54 \\
456.2359 & 0.478  & $ 0.230$ & 1.54 \\
533.0556 & 0.869  & $-0.400$ & 1.68 \\
\\
\textbf{Nd II} & \multicolumn{3}{l}{A$_\odot$(Nd) = 1.45}\\
402.1327 & 0.321 & $-0.100$ & 1.35 \\ 
405.9950 & 0.205 & $-0.520$ & 1.59 \\ 
464.5760 & 0.559 & $-0.760$ & 1.50 \\
485.9026 & 0.321 & $-0.440$ & 1.59 \\ 
495.9115 & 0.064 & $-0.800$ & 1.62 \\ 
523.4190 & 0.550 & $-0.510$ & 1.56 \\ 
527.6869 & 0.859 & $-0.440$ & 1.30 \\ 
531.9810 & 0.550 & $-0.140$ & 1.52 \\ 
\end{longtable}
\normalsize
\end{longtab}

\end{document}